\documentclass[a4paper,12pt]{article}

\usepackage[margin=1in]{geometry}
\usepackage{setspace}
\usepackage{parskip} 

\usepackage{graphicx}
\usepackage{float}
\usepackage{caption}
\usepackage{subcaption}

\usepackage{booktabs}      
\usepackage{tabularx}      
\usepackage{longtable}     
\usepackage{multirow}      
\usepackage{array}         
\usepackage{ragged2e}      
\usepackage{adjustbox}     
\usepackage{threeparttable}

\usepackage{amsmath}
\usepackage[authoryear,round]{natbib} 

\usepackage{xcolor}
\usepackage{hyperref}

\hypersetup{
    colorlinks=true,
    linkcolor=blue,
    filecolor=magenta,
    urlcolor=cyan,
    citecolor=green,
}

\newcolumntype{L}{>{\RaggedRight\arraybackslash}X} 
\title{
    \centering \Large \textbf{Signal in the Noise: Decoding the Reality of Airline Service Quality with Large Language Models}
}
\author{Ahmed Dawoud, Osama El-Shamy, Ahmed Habashy}
\date{\today}

\begin{document}

\maketitle
\thispagestyle{empty} 

\thispagestyle{empty} 

\setcounter{page}{1} 
\begin{abstract}
Traditional service quality metrics often fail to capture the nuanced drivers of passenger satisfaction hidden within unstructured online feedback. This study validates a Large Language Model (LLM) framework designed to extract granular insights from such data. Analyzing over 16,000 TripAdvisor reviews for EgyptAir and Emirates (2016–2025), the study utilizes a multi-stage pipeline to categorize 36 specific service issues. The analysis uncovers a stark "operational-perception disconnect" for EgyptAir: despite reported operational improvements, passenger satisfaction plummeted post-2022 (ratings $< 2.0$). Our approach identified specific drivers missed by conventional metrics—notably poor communication during disruptions and staff conduct—and pinpointed critical sentiment erosion in key tourism markets. These findings confirm the framework’s efficacy as a powerful diagnostic tool, surpassing traditional surveys by transforming unstructured passenger voices into actionable strategic intelligence for the airline and tourism sectors.
\end{abstract}
\newpage

\section{Introduction}
In the fiercely competitive global air travel market, understanding and improving airline service quality is paramount. It shapes brand perception, drives passenger loyalty, and holds significant economic weight, particularly for national carriers like EgyptAir, which serve as critical conduits for tourism – a vital sector for the national economy \citep{oconnell2011}. Traditionally, airlines have relied on operational metrics (for example, On-Time Performance, safety ratings) and structured passenger surveys (often based on frameworks like SERVQUAL) to gauge performance \citep{elsakty2014}. Although valuable, these methods provide a high-level view and often struggle to capture the nuanced, real-time passenger experience. Operational data reveal what happened (e.g., a flight was delayed), but not why passengers felt a certain way about it, while surveys are limited by predefined questions and may not uncover unexpected or granular issues.

However, the digital era presents an immense, largely untapped resource: vast quantities of unstructured text data in the form of authentic passenger reviews on platforms like TripAdvisor. This data offers rich and spontaneous narratives detailing lived experiences, containing potentially deep insights into the specific drivers of satisfaction and dissatisfaction. The core challenge lies in effectively processing this massive, multilingual, and often colloquial text to extract consistent, actionable intelligence at scale. Previous text analysis attempts in this domain, while useful, often lack the sophistication to fully interpret semantic nuance, handle diverse phrasing of the same issue, or operate efficiently across large multilingual datasets \citep{brochado2019}.

This gap highlights the rationale for exploring more advanced analytical techniques. Emerging Artificial Intelligence, specifically Large Language Models (LLMs), possess sophisticated natural language processing capabilities that promise to overcome these limitations. LLMs can interpret context, understand sentiment, identify specific topics, and process diverse languages with unprecedented accuracy, potentially transforming raw reviews into structured, quantifiable insights \citep{haleem2022}.

Therefore, this paper introduces and evaluates an LLM-powered analytical framework as a pilot study to demonstrate its effectiveness in unlocking deep passenger insights from online reviews. The primary goal is not merely to analyze airline performance, but to validate a methodology capable of moving beyond traditional metrics and revealing the granular 'why' behind client perceptions. To showcase and test the capabilities of this framework, we conducted a comparative analysis of EgyptAir, Egypt's flag carrier of significant national importance, and Emirates Airlines, widely regarded as an industry benchmark for service. By applying the framework to a large dataset of TripAdvisor reviews (2016-2025), we aim to demonstrate its power in identifying specific service pain points, tracking perception trends, and revealing critical insights that can inform strategic decision making for service improvement.

\section{Objectives}
The primary objective of this pilot study is to establish and validate an LLM-powered analytical framework capable of transforming unstructured, multilingual online reviews into structured, quantifiable intelligence. By applying this framework to a comparative dataset of EgyptAir and Emirates, the study aims to achieve three specific goals:
\begin{enumerate}
    \item \textbf{Methodological Validation:} To demonstrate the efficacy of LLMs in automating the extraction of granular service issues and sentiment patterns from raw text, overcoming the scalability limitations of manual content analysis.
    \item \textbf{Diagnostic Depth:} To uncover deep, actionable insights regarding passenger perception—specifically the divergence between operational metrics and experiential reality—that traditional quantitative surveys often miss.
    \item \textbf{Strategic Application:} To prove the framework’s practical value as a diagnostic tool for identifying critical service gaps and informing data-driven decision-making within the airline and tourism sectors.
\end{enumerate}

\section{Literature Review}
Research assessing EgyptAir's service quality over the past 15 years predominantly employed quantitative survey-based methodologies, often grounded in established frameworks like SERVQUAL or its airline-specific adaptations (e.g., AIRQUAL). Studies frequently used structured questionnaires with Likert scales, administered to samples typically ranging from 200 to 400 passengers, to measure perceptions across dimensions such as tangibles, reliability, responsiveness, assurance, and empathy \citep{elsakty2014, hussain2015}. Statistical techniques such as descriptive analysis, correlation, and regression were standard tools to quantify the relationship between these predefined service dimensions and outcomes such as passenger satisfaction or loyalty. These approaches provided valuable structured data, allowing statistical comparisons and identifying which factors are more significant, with reliability and staff-related dimensions often emerging as critical drivers \citep{elsakty2014}.

Although these survey-based methods offered quantifiable insights and allowed the testing of theoretical models, they have inherent limitations relevant to understanding the nuances of passenger experience:
\begin{itemize}
    \item \textbf{Predefined Constraints:} Surveys restrict feedback to fixed categories, potentially overlooking emergent, unexpected issues and failing to capture the rich contextual details—the `why'—behind passenger ratings \citep{brochado2019}.
    \item \textbf{Intensity Gaps:} The reliance on structured scales often fails to fully reflect the intensity or specific nature of dissatisfaction derived from lived experiences.
    \item \textbf{Respondent Factors:} These methods are susceptible to respondent bias and survey fatigue, which can skew results.
    \item \textbf{Loss of Granularity:} The aggregation of scores into dimensional averages can mask critical granular problems—frequently mentioned in passengers' own words—that do not neatly fit into standard survey categories.
\end{itemize}

Addressing the constraints inherent in survey methods, some studies have analyzed textual passenger feedback directly. However, traditional content analysis—reliant on manual coding and predefined categories—faces inherent limitations in scalability and semantic depth \citep{kirilenko2018}. It is often unable to consistently differentiate between closely related but distinct issues (e.g., distinguishing delays from poor communication about delays) or handle multilingual datasets without extensive pre-translation.

This methodological gap highlights the difficulty in systematically extracting deep, actionable insights at scale from the complex, unstructured, and multilingual passenger voice found online—underscores the need for more advanced analytical techniques, such as the LLM-powered framework proposed in this study \citep{haleem2022}, to bridge the divide between high-level metrics and granular experiential reality.

\section{Contextual Background: Comparative Performance Overview (2010--2025)}
To contextualize the passenger perception analysis, it is valuable to first review the performance trajectories of EgyptAir and Emirates across key operational and service metrics between 2010 and 2025. During this period, Emirates consistently maintained a reputation for high operational standards and service quality, often regarded as a benchmark in the industry \citep{oconnell2011}, while EgyptAir embarked on a path of significant, albeit challenging, improvement from a lower baseline.

\begin{table}[H]
    \centering
    \small
    \caption{\textbf{Operational Performance \& Safety (2010--2025)}}
    \label{tab:opsafety}
    \begin{tabularx}{\textwidth}{@{} l L L @{}}
        \toprule
        \textbf{Indicator} & \textbf{EgyptAir (2010s--2025)} & \textbf{Emirates (2010s--2025)} \\
        \midrule
        On-Time Performance &
        Major improvement mid-2010s; $\sim$70--75\% OTP by 2017, best in Africa 2017 (up from $\sim$50--60\% early 2010s). &
        Consistently high; $\sim$80\% OTP by 2018, ranked top 20 globally (second in Middle East 2018). \\
        \addlinespace
        Safety Rating (2024) &
        2/7 stars (failed fatality-free criteria). &
        7/7 stars (perfect score on audits, incident record). \\
        \addlinespace
        Fatal Accidents &
        1 major accident: 2016 Flight 804 crash, 66 fatalities. No other fatal crashes in period. &
        0 passenger fatalities: No fatal crash in history; one 2016 hull-loss (Flight 521) with all 300 onboard surviving. \\
        \addlinespace
        Baggage Mishandling &
        Historically higher mishandling; implemented tracking in 2019. 69\% drop in mishandled bags by 2020/21. &
        Advanced baggage systems; $\sim$1.3 mishandled per 1000 pax in Dubai hub (99.9\% success rate), industry leading. \\
        \bottomrule
    \end{tabularx}
\end{table}

As Table~\ref{tab:opsafety} illustrates, Emirates consistently outperformed EgyptAir in core operational reliability, including on-time performance and baggage handling efficiency, leveraging advanced infrastructure at its hub. More significantly, the safety records present a stark contrast; Emirates maintained an unblemished passenger fatality record and a top-tier 7/7 safety rating, a key factor in passenger trust. While EgyptAir is IOSA compliant and has improved safety management since 2016, the tragic Flight 804 accident in that year contributed to a substantially lower safety rating (2/7 stars), indicating that overcoming legacy safety perceptions remains a challenge despite recent operational improvements.

Beyond operational statistics, industry ratings and awards provide insight into overall service quality and perceived passenger experience, as shown in Table~\ref{tab:servicequality}:

\begin{table}[H]
    \centering
    \small
    \caption{\textbf{Service Quality \& Customer Satisfaction (2010--2025)}}
    \label{tab:servicequality}
    \begin{tabularx}{\textwidth}{@{} l L L @{}}
        \toprule
        \textbf{Metric} & \textbf{EgyptAir (2010s--2025)} & \textbf{Emirates (2010s--2025)} \\
        \midrule
        Skytrax Star Rating &
        3-Star Airline (average standard). &
        4-Star Airline (very high standard). \\
        \addlinespace
        Global Ranking &
        Not in Top 100 for much of the 2010s; ranked \#88 in 2024 after improvements. &
        Consistently in Top 10 worldwide; \#1 in 2013 \& 2016, \#3 in 2024. \\
        \addlinespace
        Major Awards &
        Most Improved Airline (Africa 2024, 2nd globally). Few international awards historically. &
        World’s Best Airline (2013, 2016); Best Inflight Entertainment (2005--2024); multiple class-specific awards. \\
        \addlinespace
        Inflight Entertainment &
        Personal screens on newer jets; limited content. &
        Industry-leading ICE system (6,500+ channels); full-service meals \& alcohol; won IFE awards 15+ years straight. \\
        \addlinespace
        Cabin Service &
        Improving but variable; business class still 2-2-2 on some aircraft; basic economy service. &
        Highly rated cabin crew; luxurious first/business (e.g., A380 bar, lie-flat suites); constant upgrades (premium economy introduced in 2020s). \\
        \addlinespace
        Passenger Feedback &
        Average customer rating $\sim$5--6/10; past complaints on older planes \& delays, now less frequent. &
        High customer rating $\sim$8/10; praised for entertainment, comfort, and professionalism; occasional inconsistency noted. \\
        \bottomrule
    \end{tabularx}
\end{table}

While these structured metrics outline general performance trajectories, they remain aggregated indicators that fail to capture the specific, experiential drivers of passenger sentiment. To decode the reality behind these numbers—particularly in the volatile post-2019 era—it is essential to analyze unstructured feedback. The following section details the methodology used to extract these granular insights.

\section{Methodology}

\subsection{Data Acquisition}
We employed Python-based web scraping to collect authentic passenger feedback from TripAdvisor between Q1 2016 and Q1 2025. The resulting dataset comprises 16,622 unstructured reviews—5,171 for EgyptAir and 11,451 for Emirates Airlines—spanning 13 distinct languages. This raw, unsolicited data forms the basis for the subsequent computational analysis.

\subsection{An LLM-Powered Framework}
We developed a multi-stage pipeline utilizing the **Clio** framework \citep{clio2024} to transform unstructured, multilingual feedback into actionable data. Unlike simple keyword matching, this framework employs semantic extraction to standardize diverse passenger narratives into a consistent taxonomy.

\textbf{Stage 1: Diagnostic Filtering} \\
To isolate actionable service failures, we filtered the dataset for ratings of 1 to 3 stars. This yielded a subset of 2,857 negative reviews (for EgyptAir) specifically containing direct expressions of dissatisfaction, serving as the basis for root cause analysis.

\textbf{Stage 2: Issue Extraction and Standardization} \\
The LLM processed reviews in their native language (spanning 13 languages) without pre-translation to preserve nuance. Using a structured prompting strategy with contextual memory, the model mapped diverse phrasing to a fixed taxonomy of 36 specific issue labels (e.g., normalizing "rude staff" and "bad attitude" to "Customer Service").
To illustrate this capability, Table \ref{tab:transformation_example} demonstrates how a single review containing multiple distinct complaints is disentangled and mapped to precise categories without context loss.

\begin{table}[h]
    \centering
    \small
    \caption{\textbf{Example of Unstructured Text to Structured Data Conversion}}
    \label{tab:transformation_example}
    
    \begin{tabularx}{\textwidth}{@{} p{5.5cm} X @{}}
        \toprule
        \textbf{Raw Passenger Review} & 
        \begin{tabular}{@{} p{5cm} l @{}} 
            \textbf{Extracted Issue} & \textbf{Category} 
        \end{tabular} \\
        \midrule
        
        \itshape \RaggedRight
        "Finally arrived in Cairo. The flight was delayed by 3 hours with zero updates from the gate agents. Once onboard, the seat would not recline, and the food was completely cold."
        & 
        \begin{tabular}[t]{@{} >{\RaggedRight}p{5cm} l @{}}
            1. Poor Communication (Delay) & Flight Disruptions \\
            \addlinespace[4pt] 
            2. Broken Seats               & In-Flight Experience \\
            \addlinespace[4pt]
            3. Poor Food Quality          & In-Flight Experience \\
        \end{tabular} 
        \\
        \bottomrule
    \end{tabularx}
\end{table}

\textbf{Stage 3: Semantic Clustering} \\
To facilitate strategic interpretation, we aggregated the 36 specific issues into eight macro-categories. To ensure accuracy in multi-issue reviews, the LLM extracted specific text snippets for each complaint before categorization, preventing thematic overlap. The final taxonomy includes:
\begin{enumerate}
    \setlength{\itemsep}{0pt}
    \setlength{\parskip}{0pt}
    \item \textbf{Airport Services:} Pre/post-flight processes (check-in, security, boarding).
    \item \textbf{Baggage Handling:} Lost, delayed, or damaged luggage issues.
    \item \textbf{Booking Issues:} Reservations, cancellations, website errors, and ticketing.
    \item \textbf{Cleanliness:} Hygiene standards within the cabin and airport facilities.
    \item \textbf{Customer Service:} Staff behavior, responsiveness, and interaction quality.
    \item \textbf{Flight Disruptions:} Schedule adherence (delays, cancellations, connections).
    \item \textbf{In-Flight Experience:} Onboard hardware, catering, entertainment, and comfort.
    \item \textbf{Safety Concerns:} Perceived risks or violations of safety protocols.
\end{enumerate}

\section{Findings: Unpacking Passenger Experience}

This section integrates metadata and deep semantic analysis to reconstruct the passenger experience. By examining operational metrics with unsolicited feedback, we expose the specific drivers of satisfaction and dissatisfaction for EgyptAir and Emirates during the post-pandemic recovery.

\subsection{The Performance Gap: Systemic Erosion vs. Resilience}
\begin{figure}[H]
    \centering
    \includegraphics[width=1\linewidth]{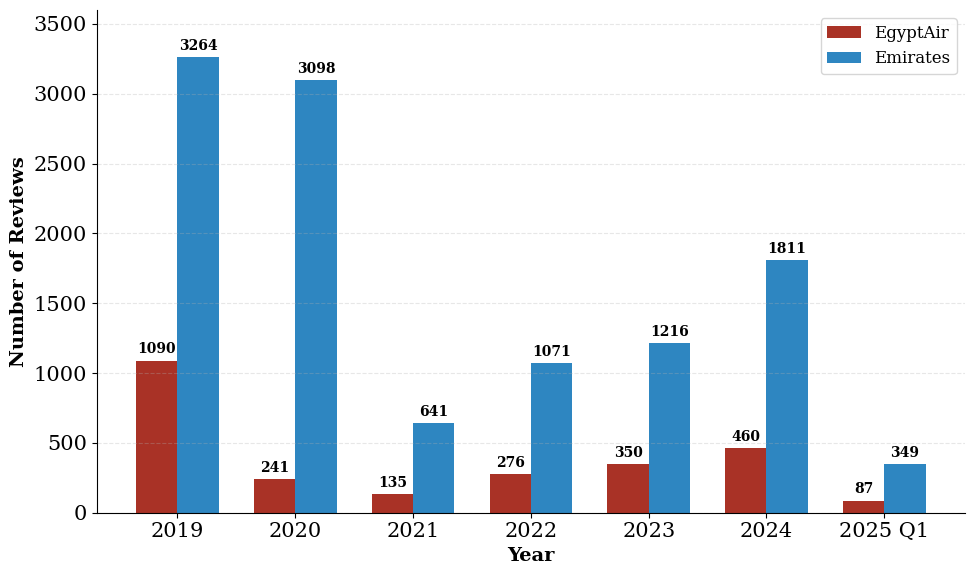}
    \caption{Review Frequency Comparison (2019-2025 Q1)}
    \label{fig:2}
\end{figure}

A comparative analysis of review volume and ratings (2019–2025) reveals two diverging trajectories. While both airlines experienced a review volume resurgence post-pandemic (Figure~\ref{fig:2}), the sentiment accompanying this return differs fundamentally.

\begin{figure}[H]
    \centering
    \includegraphics[width=1\linewidth]{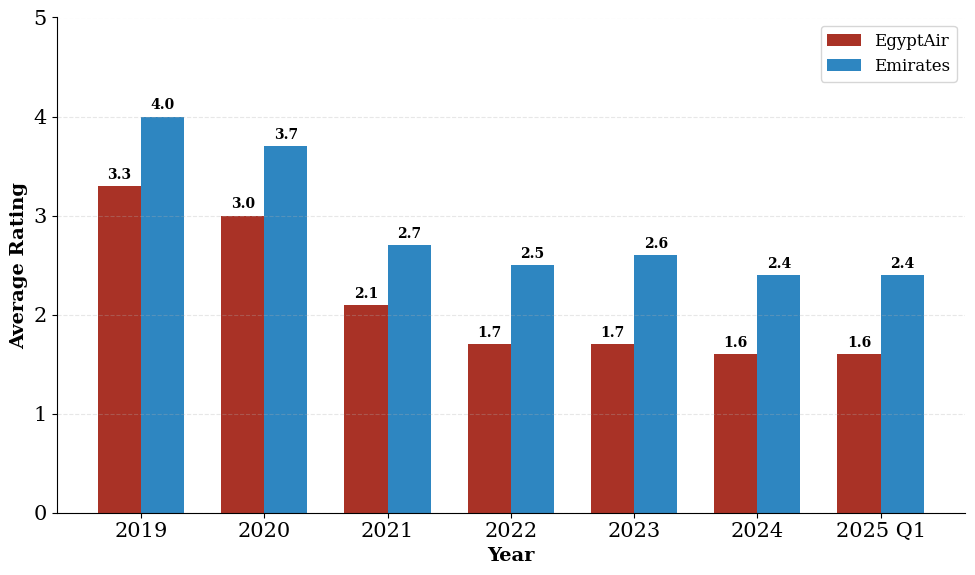}
    \caption{Average Rating Trajectory (2019-2025 Q1)}
    \label{fig:3}
\end{figure}

\textbf{The Rating Collapse:} Emirates maintained stability throughout the recovery period, with average ratings hovering above 3.5/5. In stark contrast, EgyptAir’s satisfaction levels collapsed. Starting from a modest baseline of 3.27 in 2019, ratings plummeted below the critical 2.0 threshold from 2022 onwards, settling at approximately 1.6 by 2024 (Figure~\ref{fig:3}). This severe decline occurred precisely when the airline was receiving industry accolades for operational improvements (e.g., OTP), highlighting a profound disconnect between technical performance and passenger perception.

\textbf{Systemic vs. Isolated Failure:} Deconstructing these ratings by subcategory reveals the scope of the issue.
\begin{itemize}
    \item \textbf{EgyptAir (Systemic Erosion):} Dissatisfaction is not limited to a single domain. Table~\ref{tab:detailed_issues} shows a comprehensive deterioration across all service dimensions. High-touch areas like ``Customer Service'' (271 lowest-tier ratings) and ``Value for Money'' are the primary detractors, but even hardware-related categories like ``Legroom'' and ``Cleanliness'' consistently trended toward ``Poor'' ($<2.0$) by 2025.
    \item \textbf{Emirates (Relative Stability):} While Emirates saw minor dips in ``In-flight Entertainment'' and ``Food,'' averages remained within the ``Good'' range ($>3.0$). The distribution of negative reviews shows a steep drop-off after top complaints, unlike EgyptAir’s flat distribution, indicating that Emirates' failures are isolated incidents rather than systemic flaws.
\end{itemize}

\subsection{Strategic Vulnerabilities: The Geographical Deficit}
\begin{figure}[H]
    \centering
    \includegraphics[width=1\linewidth]{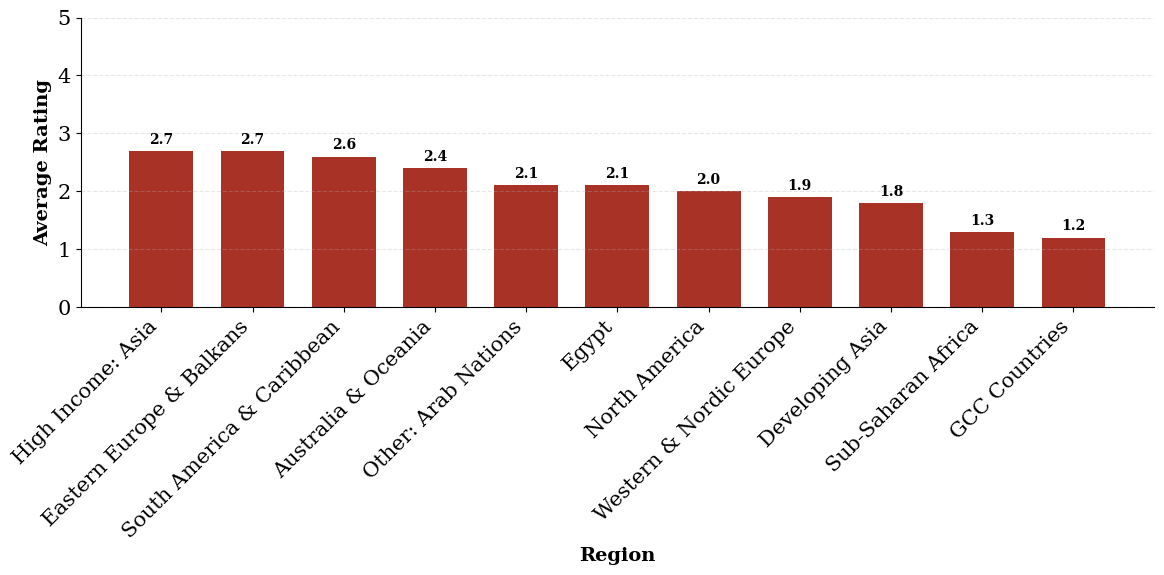}
    \caption{EgyptAir Average Rating by Origin Region (2024-2025 Q1)}
    \label{fig:egygeo}
\end{figure}

Segmenting ratings by passenger origin exposes critical strategic risks for EgyptAir, particularly in markets essential to national tourism (Figure~\ref{fig:egygeo}).

\textbf{Core Market Failure:} The lowest satisfaction levels originate from regions with the highest strategic value.
\begin{itemize}
    \item \textbf{GCC Region:} Passengers from the Gulf Cooperation Council recorded an abysmal average rating of $\sim$1.2. Given the cultural proximity and high economic value of this route, such a score represents a catastrophic failure of service alignment.
    \item \textbf{Developing Asia \& Sub-Saharan Africa:} Ratings from these massive growth markets hover between 1.4 and 2.0. This active dissatisfaction serves as a deterrent to future travel flows from regions critical to Egypt’s tourism strategy.
\end{itemize}

\begin{figure}[H]
    \centering
    \includegraphics[width=1\linewidth]{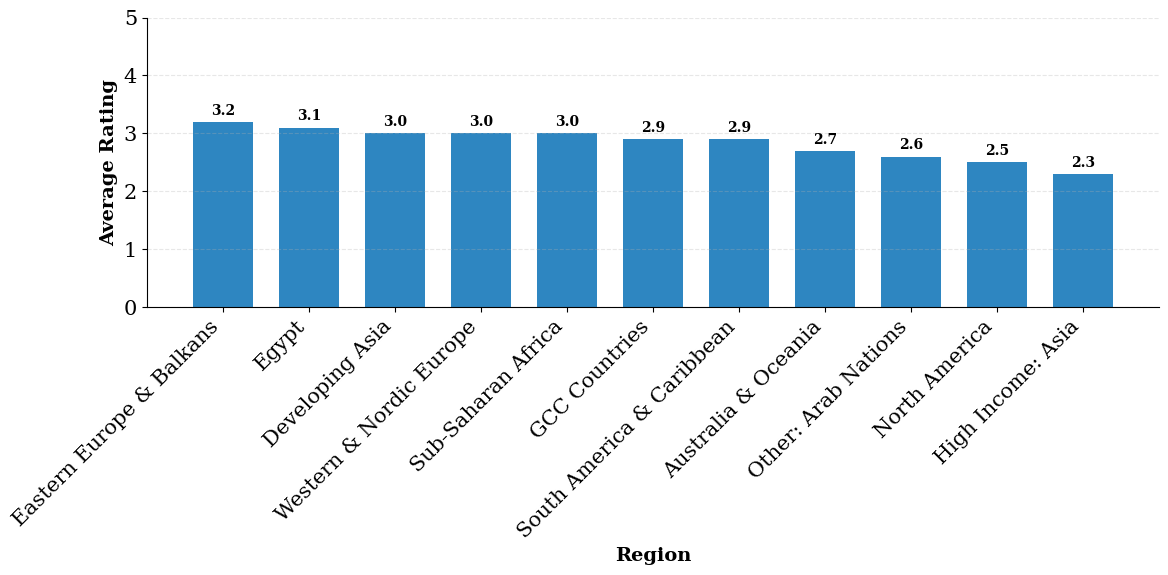}
    \caption{EgyptAir Average Rating by Origin Region (2024-2025 Q1)}
    \label{fig:emigeo}
\end{figure}

Even in relative "bright spots" like Eastern Europe, ratings barely reach 2.7, signaling mediocrity. In contrast, Emirates (Figure~\ref{fig:emigeo}) displays global consistency, with a "floor" of 2.3 and most regions clustering around 3.0–3.8. This disparity suggests EgyptAir suffers from inconsistent service delivery potentially linked to route-specific operational standards.

\subsection{Root Cause Diagnostics: From "Product" to "Process"}
While Emirates serves as a stable benchmark for comparison, this framework is essential for EgyptAir to explain why passenger satisfaction plummeted even as official operational metrics improved. Leveraging the LLM to extract 36 distinct issue types allows us to diagnose the \textit{why} behind this quantitative collapse. The analysis reveals a distinct shift in the nature of complaints from the pre-pandemic to the post-pandemic era.

The full breakdown of these specific issues is detailed below in Table~\ref{tab:detailed_issues}.

\begin{longtable}{@{} >{\bfseries}p{4cm} >{\RaggedRight}p{\dimexpr\textwidth-5.5cm\relax} r @{}}
    
    \caption{\textbf{Frequency of Service Issues by Category (2016-2025)}} \label{tab:detailed_issues} \\
    \toprule
    Main Category & Specific Issue & Count \\
    \midrule
    \endfirsthead

    \toprule
    Main Category & Specific Issue & Count \\
    \midrule
    \endhead

    \midrule
    \multicolumn{3}{r}{\textit{Continued on next page...}} \\
    \endfoot

    \bottomrule
    \endlastfoot

    
    Flight Disruptions 
     & Flight Delays/Cancellations & 690 \\
     & Poor Communication Regarding Delay & 536 \\
     & Unexplained Cancellation & 94 \\
     & Missed Connection & 92 \\
     & Excessive Flight Delay & 73 \\
     & Inadequate Pre-Flight Communication & 24 \\
     & Unclear Announcements & 1 \\
    \midrule

    Customer Service 
     & Rude Flight Attendants & 591 \\
     & Poor Customer Service & 475 \\
     & Unresponsive Crew & 346 \\
     & Lack of Assistance & 194 \\
     & Unhelpful Phone Support & 92 \\
    \midrule

    In-Flight Experience 
     & Poor Food Quality & 555 \\
     & Lack of Amenities & 442 \\
     & Uncomfortable Seating & 288 \\
     & Seat Assignment Problems & 131 \\
     & Broken Seats & 107 \\
     & Broken Entertainment System & 81 \\
     & In-Flight Experience (General) & 66 \\
     & Lack of Legroom & 12 \\
     & Seat Issues & 5 \\
     & Poor Entertainment & 1 \\
    \midrule

    Baggage Handling 
     & Lost Baggage & 335 \\
     & Damaged Baggage & 111 \\
     & Delayed Baggage & 41 \\
     & Baggage Handling Fees & 28 \\
     & Baggage Handling (General) & 2 \\
    \midrule

    Cleanliness 
     & Dirty Cabin & 280 \\
     & Unclean Restrooms & 102 \\
     & Cleanliness (General) & 2 \\
    \midrule

    Safety Concerns 
     & Lack of Safety Enforcement & 130 \\
    \midrule

    Airport Services 
     & Disorganized Boarding & 72 \\
     & Disorganized Airport Staff & 44 \\
    \midrule

    Booking Issues 
     & Difficult Booking Process & 46 \\
     & Website Issues & 37 \\

\end{longtable}

\textbf{Specific Drivers of Dissatisfaction:}
\begin{enumerate}
    \item \textbf{Flight Disruptions \& The Communication Void:} The LLM identified that ``Poor Communication Regarding Delay'' (536 mentions) is nearly as frequent as the delays themselves (690 mentions). Passengers are less frustrated by the operational irregularity than by the silence or misinformation accompanying it. This explains why technical OTP improvements (noted in Section 2) failed to improve satisfaction ratings.
    
    \item \textbf{The "Rude" Factor:} Within Customer Service, ``Rude Flight Attendants'' is the single most frequent specific complaint (591 mentions). The qualitative summaries generated by the LLM describe interactions characterized not just by inefficiency, but by \textit{hostility}—including shouting, dismissiveness, and perceived discrimination.
    
    \item \textbf{Tangible Deficiencies:} While secondary to service issues, ``Poor Food Quality'' and ``Delayed Baggage'' remain significant agitators. The persistence of baggage delay complaints suggests that while lost bag rates may be down, the speed of delivery remains a friction point.
\end{enumerate}

\begin{figure}[H]
    \centering
    \includegraphics[width=1\linewidth]{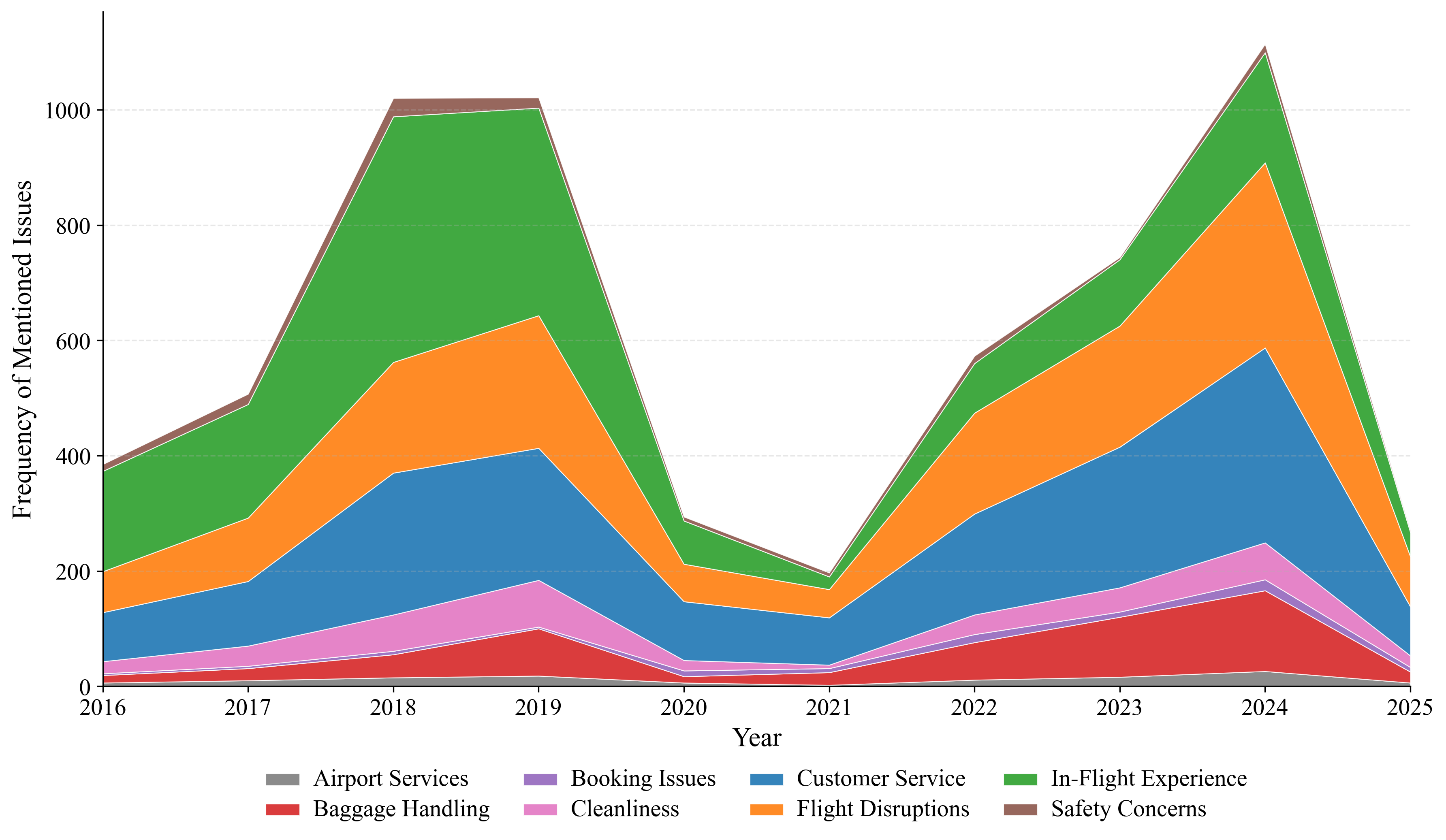}
    \caption{Temporal Evolution of Complaint Themes (2016-2024)}
    \label{fig:11}
\end{figure}

\textbf{Figure \ref{fig:11}} provides a temporal analysis of the service complaint landscape, illustrating the evolution of dissatisfaction drivers from 2016 through early 2025. The chart reveals a distinct V-shaped recovery in total complaint volume, characterized by a pandemic-induced lull (2020--2021) followed by an exponential resurgence. Critically, the post-2022 period marks a structural shift in the nature of passenger grievances. While pre-pandemic complaints were often distributed across tangible factors like "In-Flight Experience" (light blue), the recovery phase is dominated by "Flight Disruptions" (green) and "Customer Service" (dark blue). This widening of the service-related bands visually corroborates the "Operational-Perception Disconnect" identified in our analysis: as flight volume returned, the supporting infrastructure of communication and staff responsiveness failed to scale proportionally, leading to a surge in interactional rather than purely operational failures.

\subsection{Validation: LLM Framework vs. Traditional Literature}
The insights generated by our LLM-powered framework align with established literature while offering superior granularity. As detailed in Table~\ref{tab:comparison}, our findings corroborate historical issues identified in surveys—such as staff attitude and reliability \citep{elsakty2014}. However, the LLM approach distinguishes itself by uncovering the \textit{nuance} behind these categories. For instance, where traditional surveys flag "Reliability" as a problem, our framework specifically identifies that \textit{communication during delays} is a greater driver of dissatisfaction than the delay itself. Furthermore, the framework's ability to segment sentiment by geography reveals critical strategic vulnerabilities in specific tourism markets that aggregate survey data largely missed \citep{kirilenko2018}.

\begin{longtable}{@{} >{\RaggedRight}p{3.5cm} >{\RaggedRight}p{6cm} >{\RaggedRight\arraybackslash}p{6cm} @{}}
    \caption{\textbf{Comparison of LLM Framework Findings vs. Traditional Literature}} \label{tab:comparison} \\
    \toprule
    \textbf{Feature} & \textbf{LLM-Powered Framework (This Study)} & \textbf{Traditional Literature (2010–2025)} \\
    \midrule
    \endfirsthead
    \toprule
    \textbf{Feature} & \textbf{LLM-Powered Framework (This Study)} & \textbf{Traditional Literature (2010–2025)} \\
    \midrule
    \endhead
    \midrule
    \multicolumn{3}{r}{{Continued on next page}} \\
    \endfoot
    \bottomrule
    \endlastfoot

    Core Methodology & Automated LLM analysis of unsolicited, multilingual reviews (TripAdvisor). & Primarily quantitative surveys (SERVQUAL) and manual content analysis. \\
    \addlinespace
    Key Issues Identified & \textbf{Alignment:} Poor Customer Service (Rude Staff), Flight Disruptions, In-flight Comfort. & \textbf{Alignment:} Poor Staff Attitude, Reliability \citep{elsakty2014}, Tangibles gaps. \\
    \addlinespace
    Depth \& Nuance & \textbf{High:} Differentiates root causes (e.g., Delay vs. Communication); Extracts 36 specific issue types. & \textbf{Moderate:} Broad dimensions (Reliability, Responsiveness); limited "why" behind the scores. \\
    \addlinespace
    Systemic View & \textbf{Strong Evidence:} Widespread low ratings across all subcategories and regions post-2022. & \textbf{Implied:} Identifies gaps but lacks evidence of the recent, sharp divergence between ops and sentiment. \\
    \addlinespace
    Geographical Insight & \textbf{High:} Revealed catastrophic failure in strategic markets (GCC, Asia). & \textbf{Limited:} Most studies lack granular geographic segmentation. \\
    \addlinespace
    Efficiency & \textbf{High:} Automated processing of 16k+ reviews; near real-time potential. & \textbf{Low:} Surveys require significant manual effort and lag behind real-time events. \\
\end{longtable}

\section{Discussion and Conclusion: The Operational Paradox}

This study exposes a profound \textbf{chasm between operational metrics and passenger reality} for EgyptAir. While industry reports highlight improvements in On-Time Performance (OTP) and baggage handling post-pandemic, our analysis reveals that passenger satisfaction has simultaneously collapsed to historic lows. This divergence underscores a critical lesson: in the airline industry, \textbf{experiential factors outweigh technical execution.}

\subsection{Interpreting the Disconnect}
The data indicates that passengers are willing to forgive operational friction (like a delay) but rarely forgive social friction (like rudeness or silence). The dominance of complaints regarding "Poor Communication" and "Rude Flight Attendants" explains why OTP awards did not translate into higher ratings. For the passenger, a flight that arrives on time but involves hostile service is still a failure. EgyptAir’s decline is not an issue of hardware; it is an issue of \textit{software}—culture, communication, and service recovery.

\subsection{The Strategic Imperative}
The implications extend beyond the airline to the national economy. The analysis identified that EgyptAir performs worst in its most vital source markets—the GCC and Developing Asia. With average ratings near 1.2 in the Gulf, the national carrier is effectively acting as a deterrent in regions essential to Egypt’s tourism targets. This transforms the service quality deficit from an operational concern into a national strategic liability.

\subsection{Future Directions}
This study validates that Large Language Models can serve as a superior diagnostic tool for service quality \citep{haleem2022}. By moving beyond rigid survey categories to "listen" to unstructured narratives, airlines can identify the specific root causes of dissatisfaction—such as the specific nature of crew interactions or communication gaps—that traditional metrics miss. 

For EgyptAir, the path forward is clear. The data suggests that investment in new aircraft or baggage systems will yield diminishing returns on satisfaction unless accompanied by a fundamental overhaul of the service culture and communication protocols. Rebuilding trust in key markets requires shifting focus from the logistics of flying to the experience of the passenger.


\end{document}